# Universal behavior of transition temperatures Vs residual resistivity in Mn site doped La-Ca-Mn-O perovskites

L. Seetha Lakshmi , V. Sridharan[@], D. V. Natarajan , V. Sankara Sastry and T. S. Radhakrishnan[#]
Materials Science Division, Indira Gandhi Centre For Atomic Research, Kalpakkam,
Tamil Nadu 603 102, India

**Abstract**

We discuss here a comparative study of the role of local structure and/or nature of local magnetic coupling on the electrical transport properties of Mn site substituted La-Ca-Mn-O perovskites. Particular emphasis is being paid to explore the strong correlation between the insulator-metal transition ($T_{IM}$) and the residual resistivity ($\rho_o$) upon substitution. All the substitutions, irrespective of its diamagnetic or paramagnetic nature lower the transition temperatures. It is also found that, substitutions broaden the electrical as well as the magnetic transitions and increases the residual resistivity ($\rho_o$). Both the insulator-metal transition temperature ($T_{IM}$) and the para to ferromagnetic transition temperature (Tc) agree well with in a limit of ± 5 K. There exists an inverse relationship between $\rho_o$ and $T_{IM}$ in the compounds under present discussion. Best fit for $T_{IM}$ Vs $\rho_o$ could be obtained for the compounds understudy to a first order exponential decay with a functional form $T_{IM} = T_{IMO} + A\exp(-\rho_o/t)$ than that to a power law. There is a previous report wherein the similar correlation in the case of rare earth substituted manganites has been attributed to Anderson-type electron localization. The *"universal behavior"* as has been observed between $\rho_o$ and $T_{IM}$ irrespective of the electronic, magnetic and chemical nature of the substituting elements in the Mn site substituted La-Ca-Mn-O perovskites needs a rigorous theoretical investigation.

*********************************************************************
@ Corresponding Author - V. Sridharan
Email: varadu@igcar.ernet.in :
Fax: +091-04114-280081
# Present address - tsr_res@vsnl.net

**Introduction:**

Mn site substitution offers a direct probe to delineate the various factors affecting the transport properties of CMR manganites. Out of the many factors, the effective spin and the local disorder at the Mn site of mixed valent manganites found to have a dominant role in affecting its ground state properties [1-5].

RE-site substitution, in general lowers the transition temperatures and results in a phenomenal increase in the residual resistivity ($\rho_o$), which could be of several orders of magnitude. J.M.D Coey *et. al.* [6] had reported a strong correlation between $\rho_o$ and the transition temperature $T_{IM}$ in the case of single crystal, thin film and polycrystalline perovskite manganites. While explaining the concomitant magnetic and insulator-metal transitions within the framework of Anderson type of localization, L. Sheng *et. al.* [7] could bring out the correlation between $\rho_o$ and the transition temperature. To the best of our knowledge, no systematic work has been undertaken to see whether such a correlation between $\rho_o$ and $T_{IM}$ exists in the case of Mn-site substituted system as well.

With this in view, we had synthesized Mn-site substituted $La_{0.67}Ca_{0.33}MnO_3$ with paramagnetic ions (Fe and Ru) and diamagnetic ion (Ti) for different concentration of substitutions [8-9]. Substitution with other diamagnetic ions ($Al^{3+}$, $Ga^{3+}$, $Zr^{4+}$ and a certain combination of Zr and Al) were also carried out for a fixed atomic concentration of 5% to elucidate the role of chemical nature of the substitution [10]. A strong correlation between $\rho_o$ and $T_{MI}$, similar to those in RE-site substituted systems is found to exist in the Mn site substituted manganites as well.

**Experiment:**

The polycrystalline samples were synthesized by a solid-state reaction route. From the room temperature powder X-ray diffraction studies, the monophasic nature of the samples were confirmed. The crystal symmetry of the samples was determined to be orthorhombic (space group: $P_{nma}$ symmetry). The lattice parameters were estimated by Rietveld refinement program. The temperature dependence of resistivity in the temperature range 300 to 4.2 K was measured in van der Pauw geometry using silver paint for electrical contacts. ac susceptibility was measured using a home built susceptometer in the temperature interval 300K-4.2 K.

**Results and Discussion**

The compounds undergo an insulator to metal transition with a characteristic peak in the resistivity at $T=T_{IM}$. The compounds except for the Ru substituted ones, exhibit a single insulator to metal transition and an associated para-ferromagnetic transition (Tc) which compares well with $T_{IM}$. In the case of $La_{0.67}Ca_{0.33}Mn_{1-x}Fe_xO_3$ ($0 \leq x \leq 0.07$) compounds, transition temperature suppression rate ($dT_{IM}/dx$) was found to be 18 K/at.% and the ferromagnetic and metallic (FM-M) ground state remains unaltered. However, Ti substitution results in a substantial reduction of $T_{IM}$ (27 K/at.%) and FM-M ground state modifies to a cluster glass beyond x = 0.05. This is understood in terms of the competition between the weakened, on account of substitution, DE ferromagnetic and ubiquitous antiferromagnetic superexchange interactions leading to a frozen spin disorder.

In contrast to other substituted systems, $La_{0.67}Ca_{0.33}Mn_{1-x}Ru_xO_3$ ($0 \leq x \leq 0.10$) compounds exhibit two insulator- metal transitions ($T_{IM1}$ and $T_{IM2}$) (Figure: 1) with $dT_{IM}/dx$ 3 and 16 K/at. % respectively (Figure: 2). A para to ferromagnetic transition is observed in association with the former peak in the resistivity. In our previous work [9], presence of two insulator to metal transitions with different $dT_{IM}/dx$ and an increase in the lattice parameters were interpreted in terms of a mixed valent state of Ru (high spin state of $Ru^{3+}$ and $Ru^{4+}$) and a possible magnetic phase separation. The magnetoresistance measurements carried out under an applied field of 1T and 5T clearly demonstrate a substantial reduction of resistance under both the peaks in the temperature dependent electrical resistivity. The detailed analysis of the electrical and magneto transport will be published elsewhere [11].

The temperature dependence of resistivity of these compounds exhibited an up-turn passing through a minimum ($\rho_{min}$) at about 40K, a common feature in polycrystalline manganite compounds due to the "tunneling" of the spin polarized electron across the grain boundary. Increase in the value of $\rho$ at 4.2 K over the $\rho_{min}$ is typically 7% exhibiting no systematic variation either with doping concentration for a given substituent or for the chemical nature of the substituent. In the present discussion, the resistivity at 4.2 K is taken as $\rho_o$. Figure 3, 4 and 5 show the variation of $T_{IM}$ with $\rho_o$ for $La_{0.67}Ca_{0.33}Mn_{1-x}M_xO_3$ systems for M=Fe, Ti and Ru respectively. As Ru substituted system exhibited two insulator to metal transitions, $\rho_o$ is plotted against both $T_{IM1}$ (Fig. 5(a)) and $T_{IM2}$ (Fig. 5(b)). It is clearly seen that $T_{IM}$ is strongly related to $\rho_o$ and it decreases with the increase in $\rho_o$. Such a dependence has been reported for the single crystal, polycrystalline and thin films of rare -earth substituted manganites [6].

For $La_{0.67}Ca_{0.33}Mn_{1-x}M_xO_3$ (M = Fe, Ti and Ru) compounds, the $T_{MI}$ Vs $\rho_o$ curves could best fitted with the first order exponential decay of the form $T_{IM} = T_{IM0} + A \exp(-\rho_o/t)$ and are shown in figure 3, 4 and 5 respectively. Best fit could also be obtained for $La_{0.67}Ca_{0.33}Mn_{0.95}M_{0.05}O_3$ (M = Al, Ga, It, $Zr_{0.03}Al_{0.02}$, Zr, Fe and Ru) compounds [Figure 6]. Such a universal behavior had been already reported in the polycrystalline, single crystal and thin films of rare-earth substituted manganites [7]. Here we wish to emphasize that the choice of the theoretical fit is based only on a best correlation and not on any physical basis.

Based on the computational studies as well as in comparison with the experimental results on $R_{1-x}A_xMnO_3$ compounds, the strong correlation between $\rho_o$ and $T_{IM}$ is understood in terms of *Anderson type of electron localization* due to both the nonmagnetic randomness and DE spin disorder. It will be worth studying, whether we can extend similar mechanisms to the observed "universal behavior" in $T_{IM}$ Vs $\rho_o$ in the Mn site substituted manganites. We invite a rigorous theoretical reconsideration of the electrical transport properties of the Mn site substituted perovskite manganites.

**Acknowledgement:**
Authors are grateful to Ms. T. Geetha Kumary, Ms. S. Kalavathi and Dr. Y. Hariharan MSD, IGCAR for providing the low temperature experimental facilities. L. Seetha Lakshmi also thank Council of Scientific and Industrial Research, New-Delhi, India for awarding a Senior Research Fellowship.

L. Seetha Lakshmi et.al

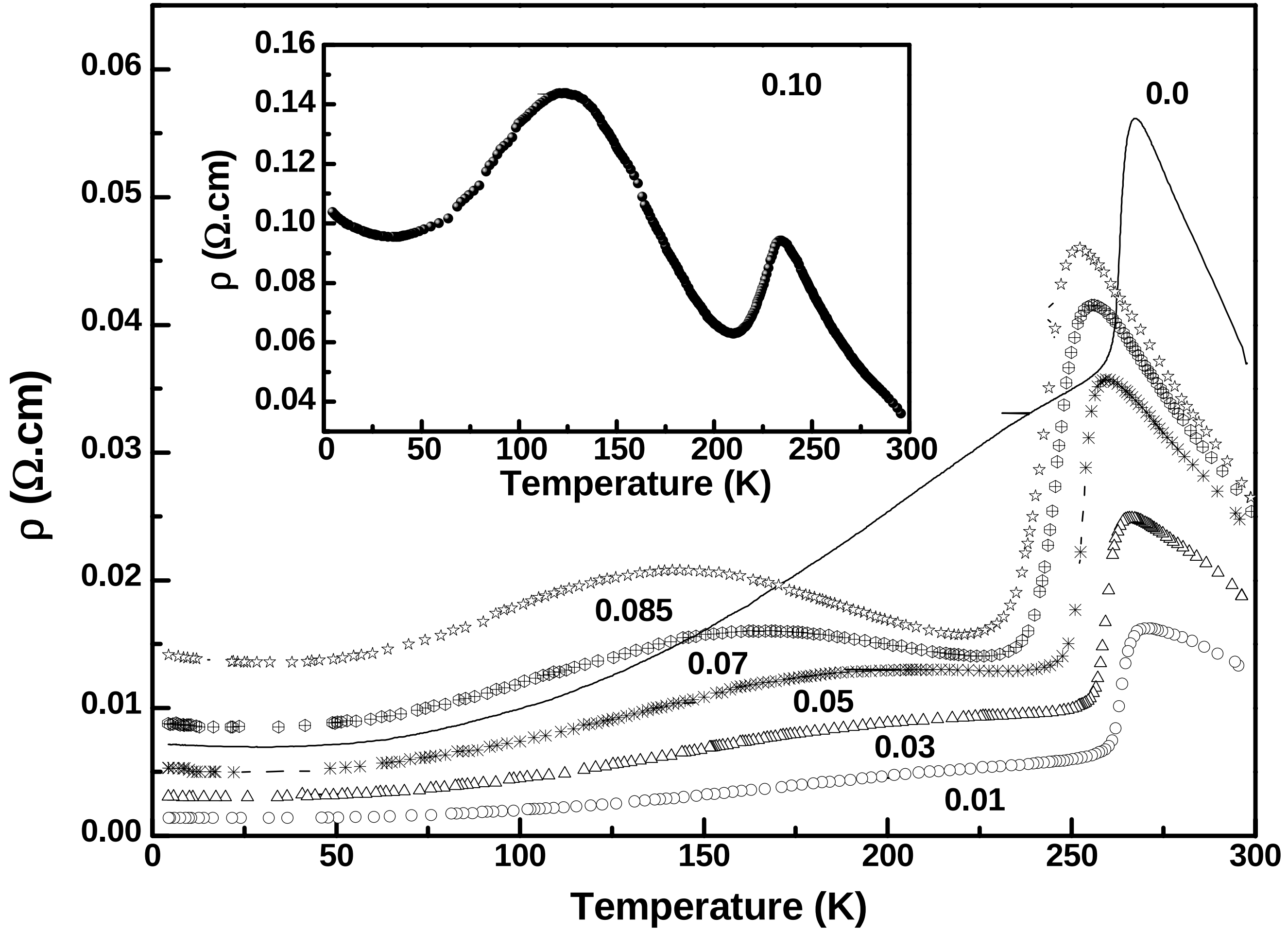

**Figure 1.** Temperature variation of resistivity of $La_{0.67}Ca_{0.33}Mn_{1-x}Ru_xO_3$ ($0 \leq x \leq 0.085$) compounds. For clarity, the temperature dependence of resistivity of $La_{0.67}Ca_{0.33}Mn_{0.90}Ru_{0.10}O_3$ compound is shown in the inset of figure 1

L.Seetha lakshmi et al

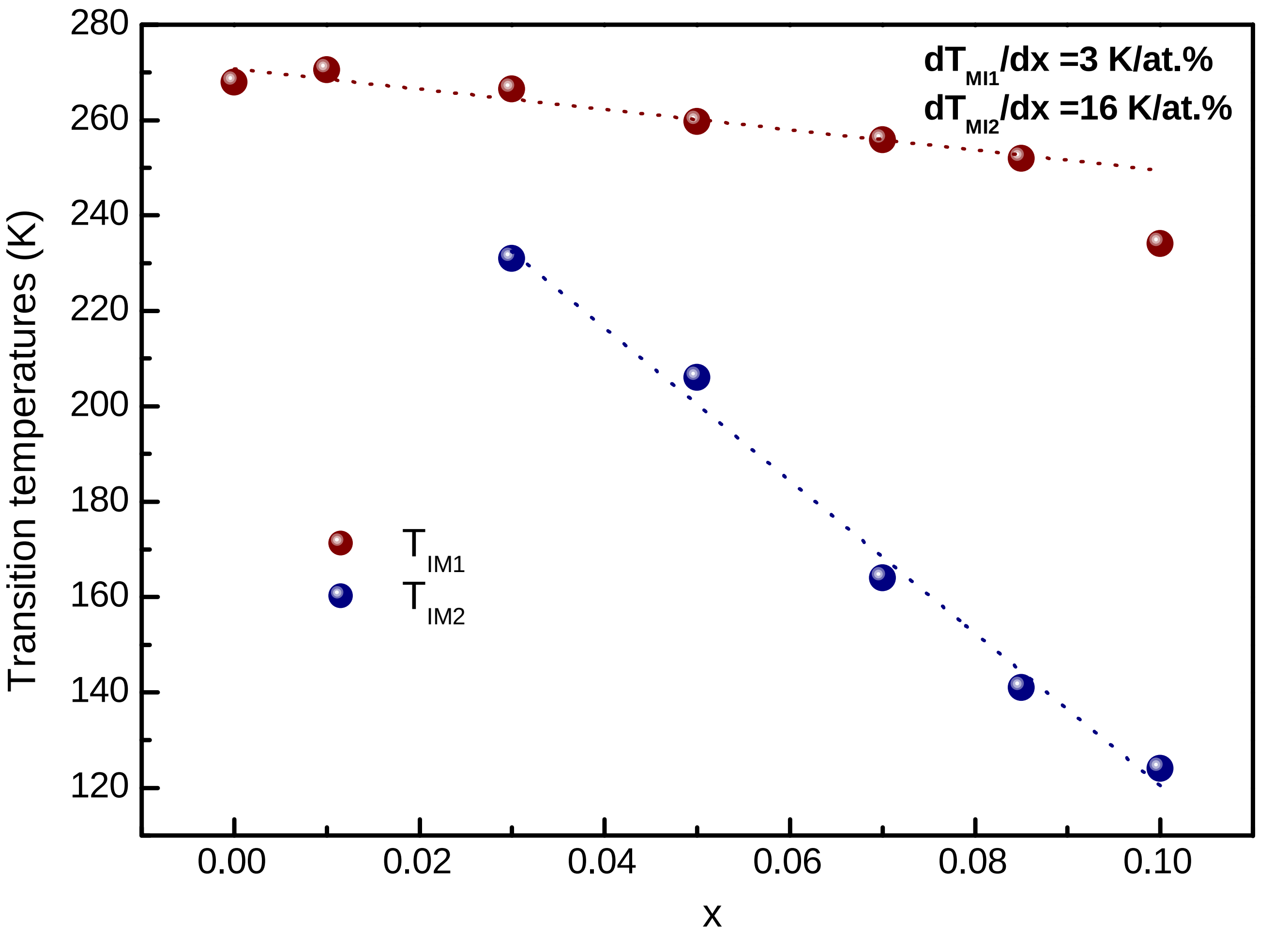

Figure :2 Variation of the insulator -metal transition temperatures ($T_{IM1}$ and $T_{IM2}$) as a function of Ru concentration of $La_{0.67}Ca_{0.33}Mn_{1-x}Ru_xO_3$ ( $0 \leq x \leq 0.10$) compounds. Solid line is the linear fit to the experimental data ( to estimate the rate of suppression of the transition temperatures )

L.Seetha Lakshmi et. al

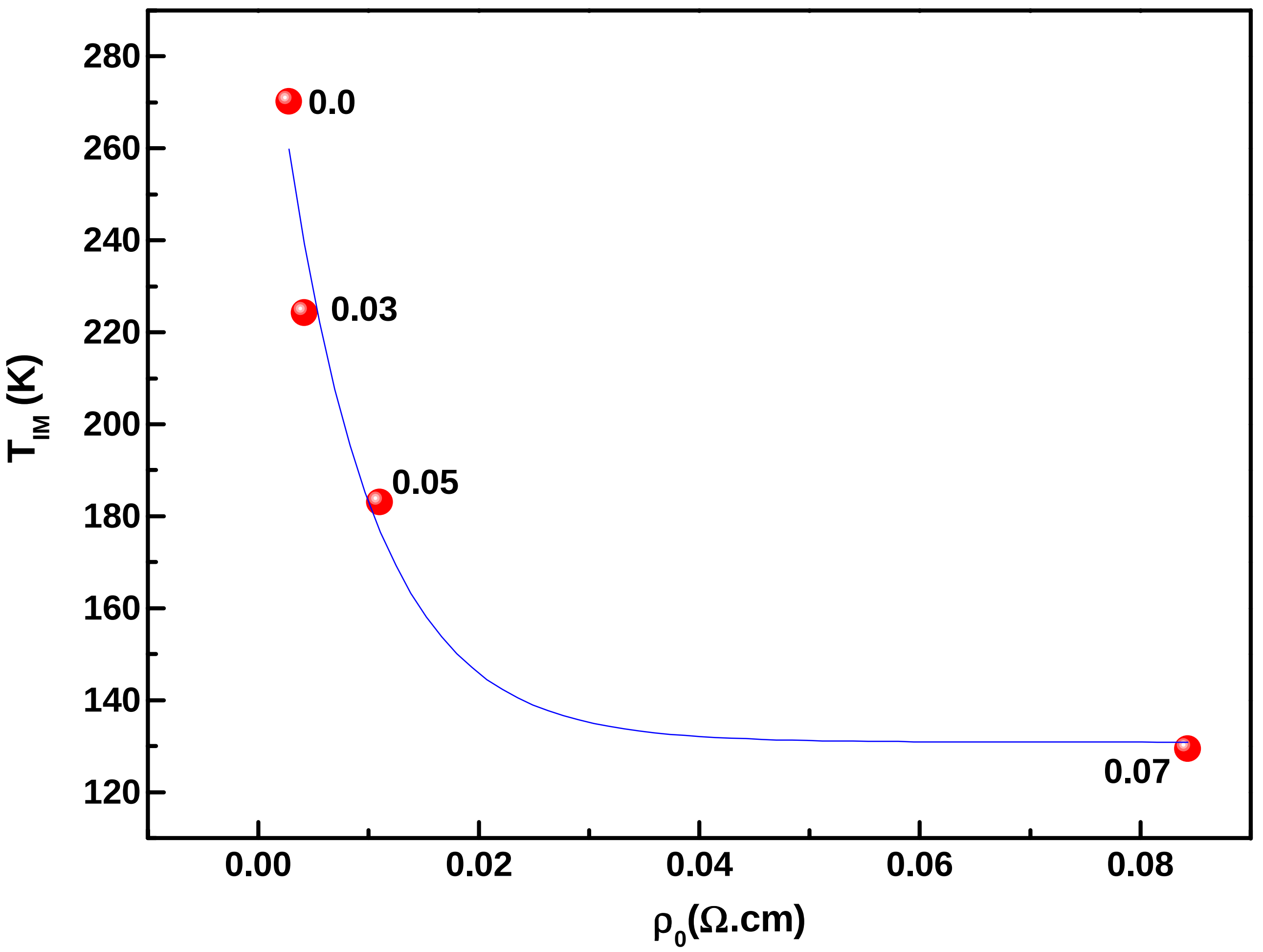

**Figure:3** Variation of insulator -metal transition temperature ($T_{IM}$) as a function of residual resistivity ($\rho_o$) of $La_{0.67}Ca_{0.33}Mn_{1-x}Fe_xO_3$ ( $0 \leq x \leq 0.07$ ) ompounds. Solid line is the best fit of the experimental data to first order exponential decay with a functional form $T_{IM} = T_{IM0} + A \exp(-\rho_o / t)$

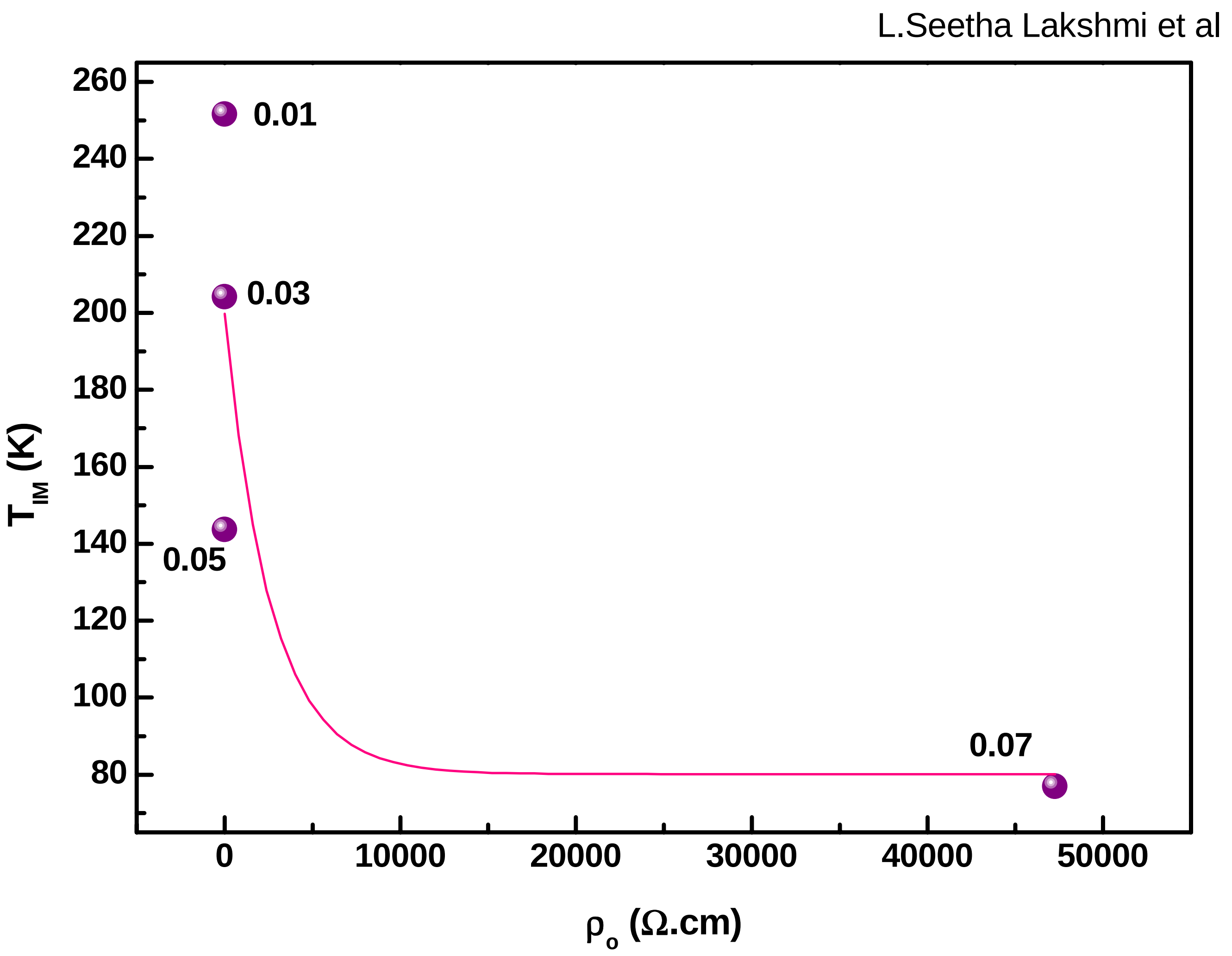

**Figure :4** Variation of insulator -metal transition temperature ($T_{IM}$) as a function of residual resistivity ($\rho_o$) of $La_{0.67}Ca_{0.33}Mn_{1-x}Ti_xO_3$ ( $0 \leq x \leq 0.10$ ) compounds. Solid line is the best fit of the experimental data to first order exponential decay with a functional form $T_{IM} = T_{IM0} + A \exp(-\rho_o / t)$

L.Seetha Lakshmi et.al

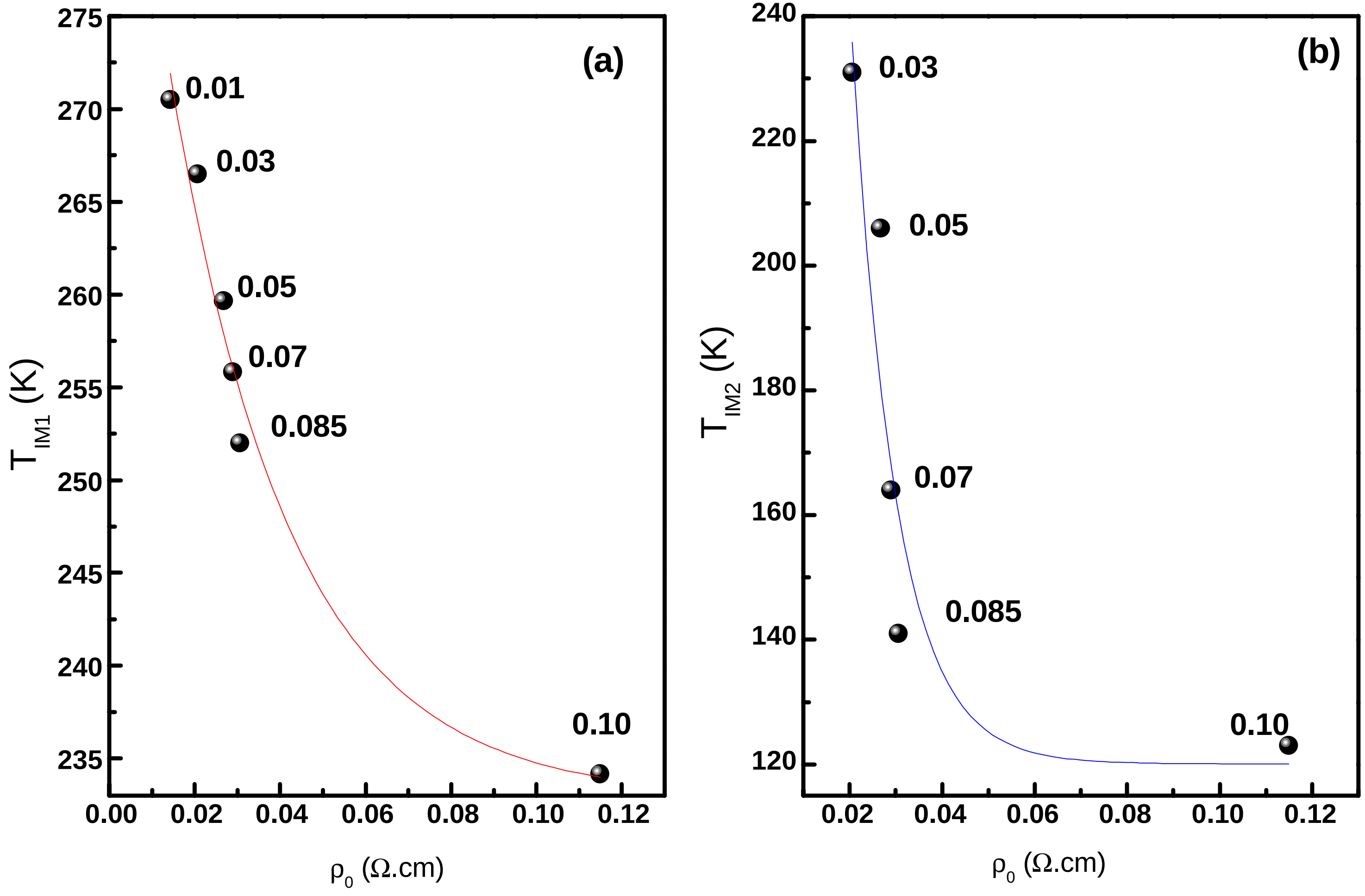

**Figure 5 (a)** Variation of insulator -metal transition temperatures ($T_{IM1}$) as a function of residual resistivity ($\rho_o$) of $La_{0.67}Ca_{0.33}Mn_{1-x}Ru_xO_3$ ($0 \leq x \leq 0.10$) compounds.

**Figure 5(b)** Variation of insulator -metal transition temperatures ($T_{IM1}$) as a function of residual resistivity ($\rho_o$) of $La_{0.67}Ca_{0.33}Mn_{1-x}Ru_xO_3$ ($0 \leq x \leq 0.10$) compounds.

Solid line in each panel is the best fit of the experimental data to first order exponential decay with a functional form $T_{IM} = T_{IM0} + A \exp(-\rho_o / t)$

L. Seetha Lakshmi et.al

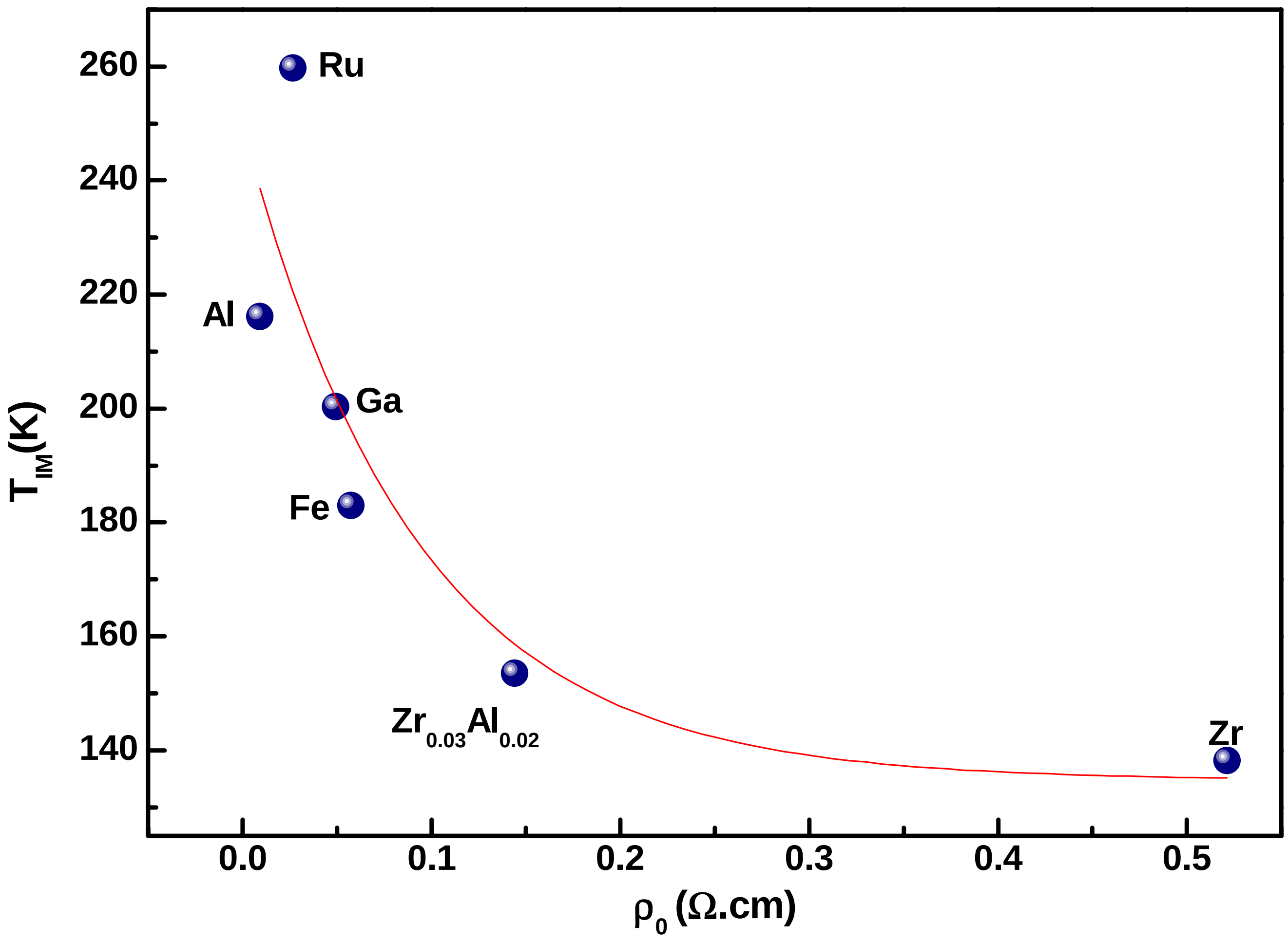

**Figure 6:** Variation of insulator-metal transition temperatures ($T_{IM}$) as a function of residual resistivity ($\rho_o$) of $La_{0.67}Ca_{0.33}Mn_{0.95}M_{0.05}O_3$ (M = Al, Ga, Ti, Zr, $Zr_{0.03}Al_{0.02}$, Ru and Fe) compounds. Solid line is the best fit of the experimental data to first order exponential decay with a functional form $T_{IM} = T_{IM0} + A \exp(-\rho_o / t)$